\documentstyle[12pt]{article}
\def\al{\alpha}

\def\ga{\gamma}
\def\de{\delta}

\def\th{\theta}

\def\ph{\phi}

\def\De{\Delta}

\def\fr#1#2{{{#1} \over {#2}}}

\def\vev#1{\langle {#1}\rangle}

\def\half{{\textstyle{1\over 2}}}
\def\frac#1#2{{\textstyle{{#1}\over {#2}}}}

\def\lsim{\mathrel{\rlap{\lower4pt\hbox{\hskip1pt$\sim$}}
    \raise1pt\hbox{$<$}}}
\def\gsim{\mathrel{\rlap{\lower4pt\hbox{\hskip1pt$\sim$}}
    \raise1pt\hbox{$>$}}}
\def\sqr#1#2{{\vcenter{\vbox{\hrule height.#2pt
         \hbox{\vrule width.#2pt height#1pt \kern#1pt
         \vrule width.#2pt}
         \hrule height.#2pt}}}}

\newcommand{\beq}{\begin{equation}}
\newcommand{\eeq}{\end{equation}}
\newcommand{\bea}{\begin{eqnarray}}
\newcommand{\eea}{\end{eqnarray}}
\newcommand{\rf}[1]{(\ref{#1})}

\renewenvironment{thebibliography}[1]
 { \rm
   \begin{list}{\arabic{enumi}.}
    {\usecounter{enumi} \setlength{\parsep}{0pt}
     \setlength{\itemsep}{3pt} \settowidth{\labelwidth}{#1.}
     \sloppy
    }}{\end{list}}

\begin{document}
\titlepage

\begin{flushright}
{COLBY-93-04\\}
{IUHET 255\\}
\end{flushright}
\vglue 1cm

\begin{center}
{{\bf RADIAL SQUEEZED STATES AND RYDBERG WAVE PACKETS
\\}
\vglue 1.0cm
{Robert Bluhm$^a$ and V. Alan Kosteleck\'y$^b$\\}
\bigskip
{\it $^a$Physics Department\\}
\medskip
{\it Colby College\\}
\medskip
{\it Waterville, ME 04901, U.S.A\\}
\bigskip
{\it $^b$Physics Department\\}
\medskip
{\it Indiana University\\}
\medskip
{\it Bloomington, IN 47405, U.S.A.\\}

}
\vglue 0.8cm

\end{center}

{\rightskip=3pc\leftskip=3pc\noindent
We outline an analytical framework for the treatment of
radial Rydberg wave packets produced by short laser pulses
in the absence of external electric and magnetic fields.
Wave packets of this type are localized in the
radial coordinates and have p-state angular distributions.
We argue that they can be described
by a particular analytical class of squeezed states,
called radial squeezed states.
For hydrogenic Rydberg atoms,
we discuss the time evolution of the
corresponding hydrogenic radial squeezed states.
They are found to undergo decoherence and collapse,
followed by fractional and full revivals.
We also present their uncertainty product
and uncertainty ratio as functions of time.
Our results show that hydrogenic radial squeezed states
provide a suitable analytical description of hydrogenic Rydberg
atoms excited by short-pulsed laser fields.

}

\vskip 1truein
\centerline{\it Published in Physical Review A,
Rapid Communications {\bf 48}, 4047 (1993)}

\vfill
\newpage

\baselineskip=20pt

Excitations of Rydberg atoms by short-pulsed laser fields
produce wave packets that are localized in the radial coordinates
\cite{ps,alber}.
These wave packets initially oscillate
with the classical keplerian period
between inner and outer
apsidal points corresponding to
those of the classical keplerian orbit
\cite{tclexps}.
They subsequently collapse and then eventually revive
almost to their original shape
\cite{revexps}.
In the interval between collapse and full revival,
subsidiary wave packets form.
These are known as fractional revivals,
and their orbital periods
are rational fractions of the classical period
\cite{revtheory}.

Depending on the excitation process and whether external
fields are present,
several different electronic orbital geometries are possible
for the quantum-mechanical motion.
For single-photon excitations without external fields,
the radial motion is localized but
the angular distribution is that of a p state.
For multi-photon excitations,
the angular distribution can be d state or higher.
If an external electric field is present at the time
of excitation,
a parabolic wave packet is formed.
This exhibits beats in the angular momentum
\cite{parabolic}.
If instead
the atomic excitation is in the presence
of a strong radiofrequency field
or crossed electric and magnetic fields,
circular atoms may be produced
\cite{circular}.
These are Rydberg atoms with high angular momentum
that are localized
in the angular coordinates.

Since Rydberg wave packets are localized and
exhibit some features of the classical motion,
a description involving some type of coherent state
\cite{kls}
would seem appropriate.
This possibility has been realized in a number of cases
for circular and elliptical quantum geometries
\cite{circs,ni},
suitable for characterizing the motion of
a hydrogenic Rydberg atom excited by
a short laser pulse in the presence of external fields.
These coherent states are either a superposition
of angular-momentum eigenstates
or eigenstates with a large value for the angular momentum.
However,
when a wave packet is produced by single-photon excitation
in the absence of external fields,
a p-state eigenfunction with $l=1$ results.

In this communication,
we show that
a particular kind of analytical squeezed state,
called a \it hydrogenic radial squeezed state, \rm
can be used to describe aspects of
a wave packet generated
by exciting a hydrogenic Rydberg atom
with a short laser pulse in the absence of external fields.
Like coherent states,
squeezed states \cite{ss}
for a given system are quantum wave functions
minimizing an operator uncertainty relation.
However,
for squeezed states,
the ratio of the operator uncertainties
typically has a different value than that of the ground state.
This means that the squeezed-state uncertainty product and ratio
display a distinct time dependence.

Direct attempts to obtain analytical squeezed states
in the variables $r$ and $p_r$
meet with a variety of obstacles.
Instead,
we have adopted an indirect method
suggested in Ref.\ \cite{ni},
in which a change of variables is made
from $r$ and $p_r$ to a new set,
$R$ and $P$,
chosen to have certain properties of harmonic
oscillator variables
and hence more amenable to analysis.
A complete derivation of the analytical form
of a general class of radial squeezed states,
also valid for Rydberg atoms other than hydrogen
(in particular the alkali-metal atoms
used in experiments),
is somewhat lengthy and
is given elsewhere
\cite{big}.
We restrict ourselves here to summarizing
the results for hydrogenic p-state excitations.
In what follows,
we work in atomic units with $\hbar = e = m_e = 1$.

The effective radial potential for a hydrogenic Rydberg
atom with angular momentum $l=1$
is $V_{\rm eff}(r) = (1-2r)/2r^2$.
For this case,
it turns out that the appropriate new quantum operator $R$
replacing the usual $r$ is
$R \equiv (2-r)/2r$,
while $P \equiv p_r$ remains unchanged.
The associated commutation relation
is $[R,P] = - i r^{-2}$,
leading to the uncertainty relation
$\De R \De P \ge \half\vev{r^{-2}}$.
We take the hydrogenic radial squeezed states
as the wave functions minimizing this uncertainty relation.
They are given by
\beq
\psi (r) = N r^{\al} e^{-\ga_0 r} e^{-i \ga_1 r}
\quad ,
\label{rss}
\eeq
where $\al$, $\ga_0$, $\ga_1$ are three real parameters
and $N$ is a normalization constant
\cite{foot}.

We fix the three parameters
$\al$, $\ga_0$, $\ga_1$
for a particular hydrogenic radial squeezed state
at time $t=0$ by matching
to the corresponding hydrogenic Rydberg wave packet at
its point of closest-to-minimum uncertainty.
This occurs at $r=r_{\rm out}$,
representing the outer apsidal point
\cite{ps}.
In the present case,
we assume that the laser excites a range of energy
states with principal quantum numbers centered on the
value $\bar n$ and that contributions from
continuum states are negligible.
We take the outer apsidal point to be
$r_{\rm out} = \bar n^2 + \bar n \sqrt{\bar n^2 - 2}$.
This point is near the outer classical apsidal point
given by $r_1 = \bar n^2 (1+e)$,
where the eccentricity $e$ is $e= \sqrt{1-1/\bar n^2}$.
We perform the match by imposing the three conditions
\beq
\vev{p_r} = 0
\quad , \,\,\,\,
\vev{r} = r_{\rm out}
\quad , \,\,\,\,
\vev{H} = E_{\bar n}
\quad .
\label{conds}
\eeq
Here,
$E_{\bar n} = -1/{2 \bar n^2}$ is the
energy corresponding to $\bar n$.
Roughly,
these conditions mean that the hydrogenic radial
squeezed state at $t=0$
is located at the outer turning point,
has no radial velocity,
and has specified energy.
The conditions \rf{conds}
determine the three parameters $\al$, $\ga_0$, $\ga_1$
in terms of  $\bar n$ and $l\equiv 1$.
Our full three-dimensional wave packet at $t=0$
is then given by $Y_{1 0} (\th,\ph) \psi_{\bar n,l=1} (r)$.

The time evolution of hydrogenic Rydberg wave packets has been
previously studied
\cite{revtheory}.
After its formation,
a packet is expected to execute
radial oscillations with periodicity equal to the
classical orbital period $T_{\rm cl} = 2 \pi \bar n^3$.
Significant self-interference due to the gradual decoherence
of the packet is anticipated at a time
$t_{\rm int} \sim \bar n T_{\rm cl}/3\de n$,
where $\de n$ measures the range of dominant
principal quantum numbers in the initial packet.
Later, the packet should reconstitute itself
almost to its original shape.
This full revival is expected to occur at a time
$t_{\rm rev} = \fr 1 3 \bar n T_{\rm cl}$.
It should persist for several orbits,
performing oscillations with period $T_{\rm cl}$.
Between $t_{\rm int}$ and $t_{\rm rev}$,
there should be times $t_r$ when the packet is gathered into
$r$ spatially separated pieces,
oscillating with wave-function periodicity
$T_r = \fr 1 r T_{\rm cl}$.
Among the values of $t_r$ are the ones we discuss below
at $t_r = \fr 1 r t_{\rm rev}$.

The analytical form
of the hydrogenic radial squeezed states permits
relatively simple expressions
to be found for the time evolution.
As these are somewhat lengthy,
we do not provide them here
\cite{big}.
Instead,
we illustrate key features of the time evolution
of the hydrogenic radial squeezed states
by focusing on a particular example:
radial p-state excitations of hydrogen centered on $\bar n = 85$.
The associated classical orbit is one of high eccentricity,
$e \simeq 1$,
and the period is
$T_{\rm cl} = 93.3$ psec.
The outer apsidal point is
$r_{\rm out} \approx 2 \bar n ^2 \simeq 14450$ a.u.
Together with the conditions \rf{conds},
this fixes the parameters of the radial squeezed state as
$\al = 168.225$,
$\ga_0 = 0.0117465$,
and $\ga_1 = 0$.

Figure 1
shows the radial probability density
$f(r) = r^2 |\psi_{\bar n,l=1} (r)|^2$
for this radial squeezed state at various times.
The initial ($t=0$) form of the state is displayed
in Fig.\ 1a.
The packet is located around $r_{\rm out}$.
Its smooth, peaked shape reflects a small
uncertainty product,
$\De r \De p_r \simeq 0.5015$.
The initial motion of the radial squeezed state
is towards the inner turning point,
near the origin.

Figure 1b shows the packet halfway through its
first orbit, $t = \half T_{\rm cl}$.
The oscillatory nature of the probability distribution
reflects the quantum nature of the packet
near the core.
Intuitively,
the electron moves faster there and
so has greater momentum,
which in turn is reflected in the greater
local curvature of the wave function.
The uncertainty product there is
$\De r \De p_r \simeq 59.5$.

Figures 1c and 1d show the radial squeezed state
after one and two full orbits,
at $t = T_{\rm cl}$ and $t =2 T_{\rm cl}$,
respectively.
The packet is evidently repeatedly returning to the
outer turning point,
but after each successive orbit
more decoherence appears
and the quantum nature of the object
becomes more apparent.
This is reflected in the uncertainty product,
which grows from
$\De r \De p_r \simeq 1.6$
at $t = T_{\rm cl}$
to $\De r \De p_r \simeq 8.0$
at $t =2 T_{\rm cl}$.

At the time $t_{\rm int}$,
which is near $\sim 4 T_{\rm cl}$
in this example,
it is no longer possible to attribute
a single peak to the radial probability distribution.
This situation is shown in Fig.\ 1e.
The remnants of the original peak are still
visible but there are many oscillations
and other subsidiary peaks present.
The decoherence of the packet
is reflected in the uncertainty product,
which at $\De r \De p_r \simeq 45.4$
is comparable to that of the highly quantum
distribution in Fig. 1b appearing
at $t = \half T_{\rm cl}$.

At the time $t_r \simeq \fr 1 3 t_{\rm rev}$,
a fractional revival is expected to appear.
The corresponding distribution for the
radial squeezed state is shown in Fig.\ 1f.
Three spatially separated packets are
indeed visible.
Similarly, a fractional revival containing two spatially
separated packets appears at
$t_2 \simeq \fr 1 2 t_{\rm rev}$.
Its forms are shown at that time
and at half a classical period later,
at $ t\simeq\fr 1 2 t_{\rm rev} + \half T_{\rm cl}$,
in Figs.\ 1g and 1h, respectively.
These figures demonstrate that
the periodicity of the distribution in time
is indeed $\half T_{\rm cl}$,
in agreement with expectation.

Figures 1i and 1j display the radial squeezed state
near the revival time,
$ t\simeq t_{\rm rev} \simeq 2.6$ nsec,
and one classical period later.
As expected,
the distribution is once again similar to that of a
single peak
located at the outer turning point,
evolving with the classical periodicity.

Figure 2a shows the uncertainty product
$\De r \De p_r$
for the radial squeezed states as a function of time.
Again,
this plot is for hydrogen with ${\bar n} =85$.
The smallest value of the uncertainty product
in $r$ and $p_r$
is close to $\half$ at the initalization time $t=0$,
when by construction
the uncertainty product in $R$ and $P$ is minimized.
The cyclic behavior of the
uncertainty product as a function of time,
expected for a squeezed state,
is revealed.
The classical orbital periodicity is again manifest
during the first few orbits,
until the wavefunction begins to decohere.
The product $\De r \De p_r$
then remains large until the revivals form.

Figure 2b displays the uncertainty ratio
$\De r / \De p_r$ as a function of time
for the same hydrogenic radial squeezed state.
This provides a measure of the relative amount
of squeezing in the two coordinates.
Overall,
it is apparent that the squeezing in $p_r$
is greater than that in $r$ by five or six orders
of magnitude.
In fact,
initially
$\De r /\De p_r \simeq 1.2 \times 10^6$.
Halfway through the first orbit,
at $t = \half T_{\rm cl}$,
the relative squeezing reaches its minimum
of $\De r /\De p_r \simeq 8.0 \times 10^4$.
The uncertainty ratio continues to change with time,
but remains large so that $p_r$ remains squeezed.

In summary,
the characteristic features of
the hydrogenic radial squeezed states
are similar to those of Rydberg wave packets in
atoms excited by a short laser pulse without external fields.
Further details of our analysis,
including in particular the generalization
of our analytical framework
via supersymmetry-based quantum defect theory
\cite{sqdt}
to the case of alkali-metal atoms,
is presented elsewhere
\cite{big}.

\vglue 0.3cm

We enjoyed conversations with Charlie Conover
and Duncan Tate.
R.B. thanks Colby College for a Science Division Grant.
Part of this work was performed while V.A.K. was
visiting the Aspen Center for Physics.

\vfill\eject

\baselineskip=16pt

\begin{description}

\item[{\rm Fig.\ 1:  }]
Radial squeezed states of hydrogen
with ${\bar n}=85$.
The unnormalized radial probability density is plotted
as a function of the radial coordinate $r$ in a.u.\ at times
(a)  $t=0$,
(b)  $t=\fr 1 2 T_{\rm cl}$,
(c)  $t= T_{\rm cl}$,
(d)  $t= 2 T_{\rm cl}$,
(e)  $t= 4 T_{\rm cl}$,
(f)  $t \simeq \fr 1 3 t_{\rm rev}$,
(g)  $t \simeq \fr 1 2 t_{\rm rev}$,
(h)  $t \simeq \fr 1 2 t_{\rm rev} + \fr 1 2 T_{\rm cl}$,
(i)  $t \simeq t_{\rm rev}$,
(j)  $t \simeq t_{\rm rev} + T_{\rm cl}$,
where $T_{\rm cl} = 93.3$ picoseconds
and $t_{\rm rev} \simeq 2.6$ nanoseconds.

\item[{\rm Fig.\ 2:  }]
(a)  The uncertainty product $\De r \De p_r$
as a function of time in nanoseconds for a
radial squeezed state of hydrogen with
$\bar n = 85$.
(b)  The ratio of the uncertainties $\De r / \De p_r$ in
units of $10^6$ a.u.\ as a function of time in nanoseconds for a
radial squeezed state of hydrogen with
$\bar n = 85$.

\end{description}
\end{document}